\documentclass[aps,prb,twocolumn,superscriptaddress,floatfix]{revtex4-1}

\usepackage{times}
\usepackage{mathptmx}
\usepackage{amsmath}
\usepackage{txfonts}
\usepackage{bm}
\usepackage[dvips]{graphicx}
\usepackage{bmpsize}
\usepackage{tabularx}
\usepackage{natbib}
\usepackage[dvips]{hyperref}
\begin{document}

\title{Conductance quantization and shot noise of a double-layer quantum point contact}


\author{D. Terasawa}
\email[]{terasawa@hyo-med.ac.jp}
%\thanks{}
\affiliation{Department of Physics, Hyogo College of Medicine, Nishinomiya 663-8501, Japan}

\author{S. Norimoto}
%\thanks{}
\affiliation{Graduate School of Science, Department of Physics, Osaka University, Toyonaka 560-0043, Japan}

\author{T. Arakawa}
%\thanks{}
\affiliation{Graduate School of Science, Department of Physics, Osaka University, Toyonaka 560-0043, Japan}
\affiliation{Center for Spintronics Research Network, Osaka University, Toyonaka, Osaka 560-8531, Japan}

\author{M. Ferrier}
%\thanks{}
\affiliation{Graduate School of Science, Department of Physics, Osaka University, Toyonaka 560-0043, Japan}
\affiliation{Laboratoire de Physique des Solides, CNRS, Universit\'e Paris-Sud, Universit\'e Paris Saclay, 91405 Orsay Cedex, France}

\author{A. Fukuda}
%\thanks{}
\affiliation{Department of Physics, Hyogo College of Medicine, Nishinomiya 663-8501, Japan}

\author{K. Kobayashi}
%\thanks{}
\affiliation{Graduate School of Science, Department of Physics, Osaka University, Toyonaka 560-0043, Japan}
\affiliation{Institute for Physics of Intelligence and Department of Physics, The University of Tokyo, Tokyo 113-0033, Japan}

\author{Y. Hirayama}
%\thanks{}
\affiliation{Graduate School of Science and CSIS, Tohoku University, Sendai 980-8578, Japan}

\date{\today}

\begin{abstract}
The conductance quantization and shot noise below the first conductance plateau $G_0 = 2e^2/h$ are measured in a quantum point contact fabricated in a GaAs/AlGaAs tunnel-coupled double quantum well.
From the conductance measurement, we observe a clear quantized conductance plateau at $0.5G_0$ and a small minimum in the transconductance at $0.7 G_0$. 
Spectroscopic transconductance measurement reveals three maxima inside the first diamond,
thus suggesting three minima in the dispersion relation for electric subbands.
Shot noise measurement shows that the Fano factor behavior is consistent with this observation.
We propose a model that relates these features to a wavenumber directional split subband due to a strong Rashba spin--orbit interaction that is induced by the center barrier potential gradient of the double-layer sample.
%The relationship between the wavenumber and the source--drain voltage is derived; additionally, the subband edge lines as a function of the source--drain voltage are presented. 
\end{abstract}

%\pacs{73.20.Fz,73.43.-f}

%\keywords{}

\maketitle

\section{\label{intro}Introduction}

In quantum point contacts (QPCs) on two-dimensional electron gas (2DEG) systems,  nanometer-scale confinement embodies a quantum ballistic transport analogous to the transverse modes of optical waveguides.
The transverse modes or subbands are well separated in energy; thus, the conductance through a QPC becomes quantized in a unit of $G_0=2e^2/h$\,\cite{vanWees,Wharam,Thomas}, where $h$ denotes Planck's constant, $e$ the elementary charge, and the coefficient 2 expresses the spin degeneracy that is understood using the Landauer-B\"{u}ttiker formalism\,\cite{Landauer,Buttiker_LB,ButtikerImryLandauerPinhas}.
Although many theoretical studies suggested the lifted spin degeneracy state ($0.5G_0$ plateau) at zero magnetic field\,\cite{Bruus,Reilly,WangBerggren,DaulNoack,Yang1D}, this degeneracy is typically not resolved.
Instead, a small plateau appears at $0.7G_0$\,\cite{Thomas}, and has attracted considerable  interest (for a review, see \cite{Micolich}).
The Landauer-B\"{u}ttiker model has been tested by measuring shot noise, i.e., the discrete noise of the charge that is carried by particles in the probabilistic scattering process\,\cite{ButtikerPRL,Buttiker1992,Lesovik,Yurke,Martin,Blanter,Kobayashi}, in this system.
Previous shot noise measurements for QPCs on 2DEGs have contributed significantly to the elucidation of basic physics and complemented the conductance results\,\cite{Kumar,Reznikov,Roche,DiCarlo,Gershon,Hashisaka,Nakamura,Kohda}. 
Furthermore, not only the fundamental physical importance, semiconductor nanostructures with a QPC offer electronic devices that can manipulate electron charges and spins; thus, they are feasible for spintronic devices\,\cite{Wolf,Awschalom} and quantum computation\,\cite{Bennett}.
In particular, a QPC on a tunnel-coupled double layer (coupled quantum wire) is a candidate for implementing a qubit\,\cite{Bielejec,Bertoni,Ramamoorthy}.
Hitherto, several studies\,\cite{Thomas_WaveFuncMixing,Salis,Thomas_SpinProperties,Nuttinck,Fischer,Smith,Ichinokura} have been conducted that resolved the coupled wavefunction modes of double-layer systems, and the obtained information is useful for quantum engineering. %, in order to provide sufficient backgrounds for quantum engineering. 
The resolution of spin degeneracy and the generation of spin currents with only electrical controls, such as using spin--orbit interactions (SOIs)\,\cite{DattaDas,Kohda,Nichele,Quay,Kammhuber,Heedt,Srinivasan,Masuda}, remain to be addressed in future studies. 
In addition, the shot noise for tunnel-coupled QPCs should be measured, because additional degrees of freedom are expected to affect many-body interactions in the nonequilibrium regime\,\cite{Ferrier}.

In this study, we fabricated a QPC in a double-layer 2DEG of a GaAs/AlGaAs double quantum well (DQW) sample and investigated the conductance quantization in this double-layer QPC system. 
Here, we report the shot noise results when the conductance is below the first conductance plateau, $G_0$.
Previously, researchers have reported the coexistence of $0.5 G_0$ and $ \sim 0.7 G_0$ plateaus\,\cite{Nuttinck,Crook,Debray,Kohda,Das}.
Using a high mobility and low electron density double-layer sample, we observed a clear conductance plateau at $0.5 G_0$, and transconductance minima at $0.5$ and $\approx 0.7G_0$ at zero magnetic field and the lowest temperature available for the dilution refrigerator used in this experiment. 
Energy spectroscopy reveals a rich structure of subband edge (SBE) lines with three maxima inside the first SBE diamond, between the $0.5 G_0$ and $G_0$ plateaus region.
They are dependent on the magnitude and direction of the magnetic fields, and consistent with the horizontal (in wavenumber direction) subband splitting model discussed herein.
From the shot noise measurement, the Fano factor $F$, i.e., the current noise normalized to the noise of Poissonian transmission statistics, exhibits reductions at $0.5 G_0$ and  $G_0$, and a small reduction at $0.7 G_0$.
In addition, we observe a difference in $F$ with regard to the positive and negative biases that further suggests an SOI dispersion with Zeeman splitting.
We hypothesize that this splitting is caused by the Rashba SOI\,\cite{Rashba} that is induced by a strong potential gradient of the center barrier and the high mobility of the sample.
%Moreover, the relationship between the wavenumber and the nonequilibrium source--drain bias is clarified in this article. 
%From this relationship, the SBE lines are derived as a function of the source--drain bias. 
This study would invoke further investigations for spin-related physics and a quasiparticle’s charge in the double-layer system.

The remainder of this paper is structured as follows: In Sec.\,\ref{experiment}, we describe the sample of this experiment (\ref{Sample}), the experimental setups for conductance measurements (\ref{ConductanceSetup}), and the shot noise measurements (\ref{ShotNoiseSetup}).
In Sec.\,\ref{Results}, we present the experimental results on the conductance measurement (\ref{ConductanceResult}) and shot noise measurement (\ref{ShotNoiseResult}). 
A discussion is presented in Sec.\,\ref{discussion}. 
After calculating the wavefunctions in the DQW at the QPC (\ref{Simulation}), we discuss the effect of the SOI for the conductance and shot noise (\ref{SOISplitting}).
We present the conclusions in Sec.\,\ref{conclusion}.
Future perspectives are presented briefly in this section.

\section{Experiment}
\label{experiment}

\subsection{Sample preparation}
\label{Sample}

\begin{figure}  
\includegraphics[width=0.86\linewidth]{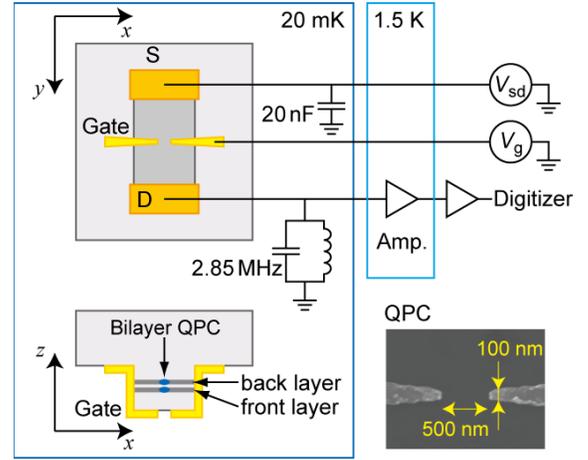}
\caption{\label{fig_sample}(Color online)\,Schematics of the sample and current noise measurement setup. The sample is placed upside down on the cold finger of the mixing chamber, as shown in the horizontal view in the bottom panel. Right inset: Scanning electron microscopy image of the split gates.  }
\end{figure}

The sample used in this study was fabricated on a DQW heterostructure grown by molecular beam epitaxy on a GaAs (100) surface in the NTT Basic Research Laboratories.
The wafer comprises two 20-nm-wide GaAs quantum wells separated by a 3-nm-wide AlAs barrier layer;  thus, the center-to-center distance $d$ is $d = 23$\,nm.
The DQW was located 600\,nm below the surface, and was doped from both sides using $1 \times 10^{12}$\,cm$^{-2}$ Si $\delta-$dopings 200\, nm away from both layers. 
The energy gap between the DQW symmetric and anti-symmetric states $\Delta_{\rm SAS}$ was measured to be 0.29\,meV through the analysis of Shubnikov de-Haas (SdH) oscillation at low magnetic fields (see Appendix \ref{SdH}).
The total electron density is $1.20 \times 10^{11}$ cm$^{-2}$, with $0.64 \times 10^{11}$ cm$^{-2}$ in the symmetric state and $0.56 \times 10^{11}$ cm$^{-2}$ in the anti-symmetric state.
The sample was processed in a shape of a standard Hall bar of width 50\,$\mu$m
and four voltage probes separated by 180\,$\mu$m (see Fig.\,\ref{fig_sample}). Two of the probes were used in this experiment.
Ohmic contacts were created using AuGe/Ni metals.
They were contacted with both layers simultaneously.
Subsequently, a pair of split gates of width 500\,nm and length 100\,nm was created, under which a coupled double-layer QPC was formed.
The scanning electron microscopy image of the split gates is shown in Fig.\,\ref{fig_sample}.
In this setup, the conductance and current noise are the results of the transport measurement through this QPC.
The low temperature electron mobility is as high as $\approx 2.5 \times 10^6 $\,cm$^2$/(Vs), given the low electron density in the DQW.
This value provides the mean free path of $ \approx 14$\,$\mu$m and the momentum relaxation time of $\approx 95$\,ps from the Drude model.
The sample was mounted on the cold finger of the mixing chamber of a dilution refrigerator with a base temperature of 20\,mK.
We determine the $x$, $y$, and $z$-directions with regard to the current flow direction through the QPC and the 2DEG plane: the $x$-direction is perpendicular to the current and in-plane to the 2DEG; the $y$-direction is parallel along the current and in-plane to the 2DEG; the $z$-direction is perpendicular to the 2DEG.  
Magnetic fields ${\bm B}=(B_x,B_y,B_z)$ were applied using a vector magnet, with maximum fields of $B_x = 3, B_y=1,$ and $B_z=8$\,T.
We use $B=|{\bm B}|$ as the magnitude of the total magnetic fields; thus, $B=0$\,T represents $B_x=B_y=B_z=0$\,T.

%%% FIGURE 1 %%%

\subsection{Conductance Measurement}
\label{ConductanceSetup}

We measured the two-terminal differential conductance $G = dI_{\rm sd}/dV_{\rm sd}$ ($I_{\rm sd}$ and $V_{\rm sd}$ denote the source--drain current and voltage, respectively) and the transconductance $dG/dV_{\rm g}$ ( $V_{\rm g}$ denotes the gate voltage applied to the split gates) simultaneously, using two lock-in amplifiers.
First, $G$ was measured using a standard lock-in technique with a frequency of 387\,Hz and amplitude of $V_{\rm sd}^{\rm ac} =10$\,$\mu$V r.m.s.; simultaneously, a small ac gate modulation $V_{\rm g}^{\rm ac}=4$\,mV r.m.s. was applied through the second lock-in amplifier with a frequency of 13\,Hz.
The output signal of the first lock-in amplifier, which includes the ac modulation signal from $V_{\rm g}^{\rm ac}$, was input to the second lock-in amplifier, whose ac modulation was referenced by itself.
This method allows us to measure the transconductance directly; therefore, it is sensitive enough to detect a small change in the transconductance.
A dc gate voltage $V_{\rm g}^{\rm dc}$ was also applied to the sample; thus, the total voltage applied to the split gate $V_{\rm g}$ is $V_{\rm g}=V_{\rm g}^{\rm dc} + V_{\rm g}^{\rm ac}$. 
In addition, a dc voltage $V_{\rm sd}^{\rm S}$ was applied to the source to cancel the voltage arising from the Seebeck effect because the drain was grounded at the mixing chamber, and dc voltage $V_{\rm sd}^{\rm dc}$ was applied to the source electrode. Thus, the total voltage applied to the source $V_{\rm sd}$ was $V_{\rm sd} =  V_{\rm sd}^{\rm ac} + V_{\rm sd}^{\rm dc} - V_{\rm sd}^{\rm S}$. 
For practical use in graphs and image plots, we ignored the ac component of $V_{\rm g}$ and $V_{\rm sd}$.

\subsection{Shot Noise Measurement}
\label{ShotNoiseSetup}

%%% TABLE 1 %%%
\begin{table}[t]%[H] add [H] placement to break table across pages
\caption{\label{typicalvalue}Typical values of parameters for noise measurement.}
\begin{ruledtabular}
\begin{tabular}{ccccc}
$A$  & $Z_0$ ($\Omega$) & $C$ (pF) & $S_{\rm V}^{\rm out}$ (V$^2/$Hz) & $S_{\rm I}^{\rm out}$ (A$^2/$Hz) \\ \hline
$8.7 \times 10^5$ & $6.1 \times 10^4$ & $1.0\times 10^{2}$ & $1.3 \times 10^{-19}$ & $6.0 \times 10^{-28}$ \\
\end{tabular}
\end{ruledtabular}
\end{table}

The current noise, i.e., the current fluctuation around its average, was measured at 300\,mK following Refs.\,\cite{Nishihara,Arakawa_APL,Muro}.
The voltage fluctuation generated in the parallel circuit of the sample and a 2.85-MHz {\it LC} resonator was measured as an output signal of a homemade cryogenic amplifier\,\cite{Arakawa_APL} at a 1\,K pot and a room-temperature amplifier, as shown schematically in Fig.\,\ref{fig_sample}.
Subsequently, the time-domain noise signal acquired by a digitizer was converted to a power spectrum through fast Fourier transform (FFT).
The current spectral density $S_{\rm I}$ was obtained by fitting the resonance peak $P_0$ that was described as a function of the sample differential resistance $R_{\rm d} = 1/G$ at a finite $V_{\rm sd}$,
\begin{equation}
P_0 = A \left[S_{\rm V}^{\rm out}+\left(\frac{Z_0R_{\rm d}}{Z_0 + R_{\rm d}}\right)^2(S_{\rm I}^{\rm out}+S_{\rm I})\right],  \label{EqP_0}
\end{equation}
where $A$ denotes the total gain of the cold and room-temperature amplifiers, $Z_0$ denotes the impedance of the {\it LC} resonance circuit, and $S_{\rm V}^{\rm out}$ and $S_{\rm I}^{\rm out}$ denote the current and voltage noise of the amplifier, respectively.
After a series of careful calibration procedures, we obtained the parameters as shown in Eq. (\ref{EqP_0}).
Their typical values are tabulated in TABLE \ref{typicalvalue}.

For a finite temperature, $S_{\rm I}$ is described by the following equation\,\cite{Blanter}:
\begin{equation}
S_{\rm I}=\frac{2F}{R_d}\left[ eV\coth \left(\frac{eV}{2k_{\rm B}T_{\rm e}} \right) -2k_{\rm B}T_{\rm e} \right] +  \frac{4k_{\rm B}T_{\rm e}}{R_{\rm d}}     \label{EqS_I}
\end{equation}
where $T_e$ denotes the electron temperature and $F$ denotes the Fano factor.
For high bias region ($|eV| >2k_{\rm B}T_{\rm e}$), the equation above becomes simpler; $S_{\rm I}$ behaves linearly on $\langle I_{\rm sd} \rangle $ as
\begin{equation}
S_{\rm I} =2eF \langle I_{\rm sd} \rangle.  \label{EqS_ILinear}
\end{equation}
We evaluated the Fano factor using this simpler form as it yielded more reliable values\,\cite{Muro}.

\section{Results}
\label{Results}

\subsection{\label{ConductanceResult}Results of the Conductance Measurement}

%%% FIGURE 2 G at 2K and 35 mK %%%
\begin{figure} 
\includegraphics[width=1\linewidth]{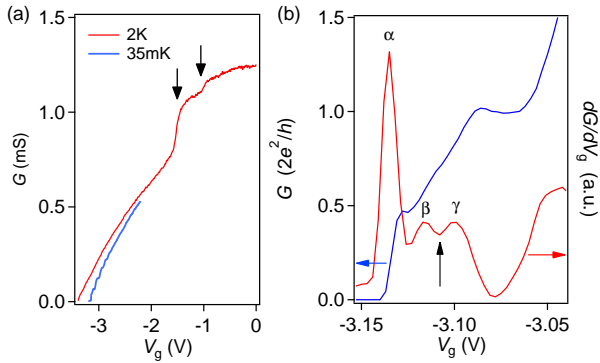}
\caption{\label{fig_G2K35mK}(Color online)\,(a) $G$ as a function of $V_{\rm g}$ at 2\,K and 35\,mK at zero magnetic field $B=0$\,T. (b) $G$ (left axis, in the unit of $G_0=2e^2/h$) and $dG/dV_{\rm g}$ (right axis, in arbitrary unit) as a function of $V_{\rm g}$.  }
\end{figure}

%%% FIGURE 3 %%%
\begin{figure*} 
\includegraphics[width=0.86\linewidth]{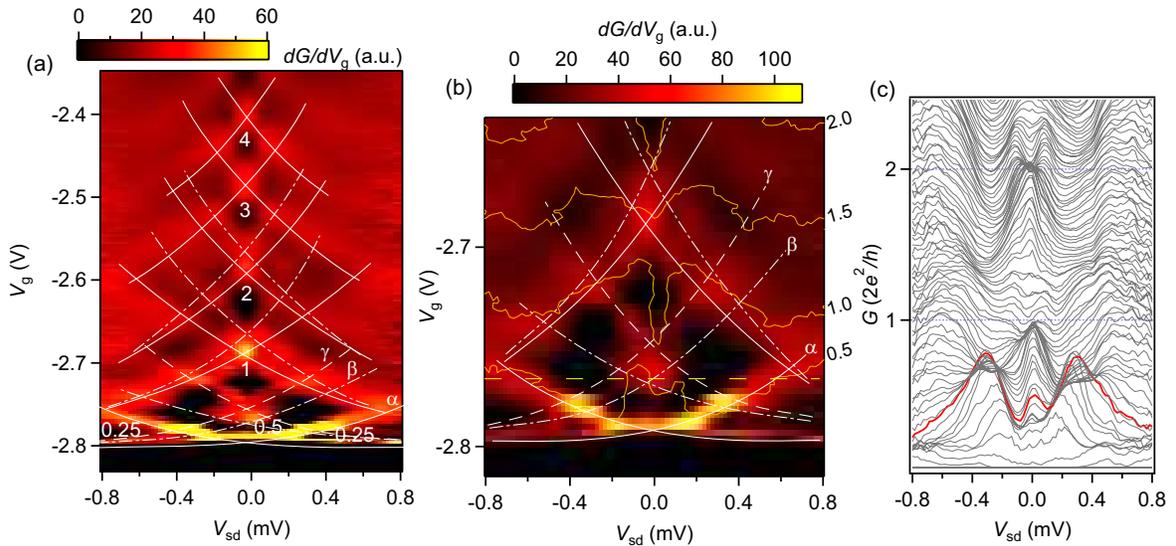}
\caption{\label{fig_dGdVgB0T}(Color online)\,(a) Image plot of $dG/dV{\rm g}$ as a function of $V_{\rm sd}$ and $V_{\rm g}$  at $T=20$\,mK and $B=0$\,T with primary SBE lines (solid lines) and with full SBE splitting lines (dash-dotted lines and a broken line), which were drawn based on the $dG/dV_{\rm g}$ maxima. The numbers express the plateau values in the units of $G_0 = 2e^2/h$. (b) Enlarged image plot of $dG/dV_{\rm g}$ with contours of $G$ in the units of $G_0$ (indicated by the slanted numbers near the right axis). The line profile at the dashed yellow line is shown later in Fig.\,\ref{fig_dGdVgBxCf} (d). (c) $G$ in the units of $G_0$ as a function of $V_{\rm sd}$ for various $V_{\rm g}$. }
\end{figure*}

%%% FIGURE 4 %%%
\begin{figure}[b] 
\includegraphics[width=1\linewidth]{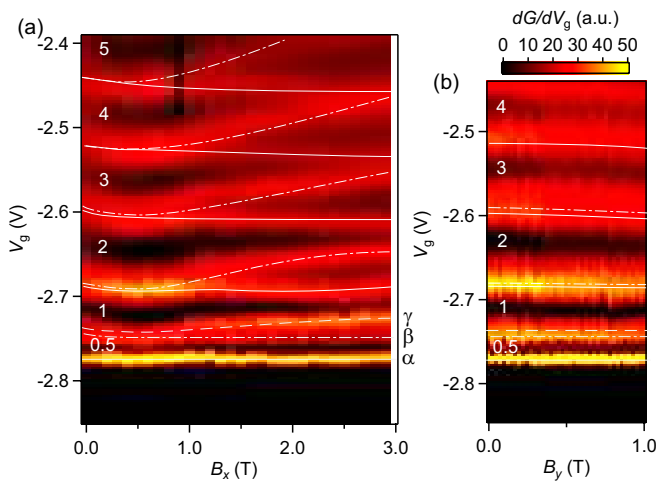}
\caption{\label{fig_dGdVg_VgvsBxBy}(Color online)\,Image plots of $dG/dV{\rm g}$ as a function of (a) $B_x$ and $V_{\rm g}$ at $B_y=0$\,T, and (b) $B_y$ and $V_{\rm g}$ at $B_x=0$\,T. }
\end{figure}

Figure \ref{fig_G2K35mK} (a) shows $G$ as a function of $V_{\rm g}$ at 2\,K and 35\,mK. 
Reflecting the property of double-layer systems at 2\,K, $G$ drops twice at $V_{\rm g} \approx -1.0$ and $-1.5$\,V (indicated by the downward arrows), corresponding to the depletion of the front and back 2DEGs under the split gate, respectively.
Then at 35\,mK, several conductance plateaus are observed for $V_{\rm g} < -2.8$\,V before the channel is pinched off at $V_{\rm g} = -3.14$\,V. 
Figure \ref{fig_G2K35mK} (b) shows detailed structures of $G$ and $dG/dV_{\rm g}$ for $G < 1.5G_0$. 
The resistances of the leads and at the contacts are subtracted accordingly.
We observe a clear $ 0.5 G_0$ plateau in $G$ and a local minimum in the $dG/dV_{\rm g}$ with a small plateau around $0.7 G_0$ (indicated by the upper arrow). 
The simultaneous observation of these two features for $B=0$\,T has been reported in several experiments\,\cite{Nuttinck,Crook,Kohda,Rokhinson,Das}. 
To the best of our knowledge, however, this has never been observed in a double-layer system before.
To supplement the explanation, unlike the typical so-called ``0.7 anomaly'' in that a relatively higher temperature is required to observe a plateau-like feature\,\cite{Thomas}, this minimum in $dG/dV_{\rm g}$ is clearly observed at extremely low temperatures such as $T \leq 35$\,mK, indicating that it originates in a ground state.
In addition, a 0.7 plateau is evolved into a clear 0.5 plateau by changing the electron density\,\cite{Thomas_SpinProperties,Nuttinck,Reilly}, or by increasing the in-plane magnetic field parallel to the channel\,\cite{Thomas}. 
Therefore, the concurrent observation of 0.5 and 0.7 plateaus is rather unusual.
Physically, the peaks observed in $dG/dV_{\rm g}$ imply that the Fermi energy crosses the SBEs.
In Fig. \ref{fig_G2K35mK} (b), three peaks are shown between the $G = 0$ and $G_0$ regions, suggesting that the Fermi energy crosses three SBEs in this region.
We name these three peaks as $\alpha$, $\beta$ and $\gamma$ from low to high $V_{\rm g}$.

%%% FIGURE 5 %%%
\begin{figure*} 
\includegraphics[width=1\linewidth]{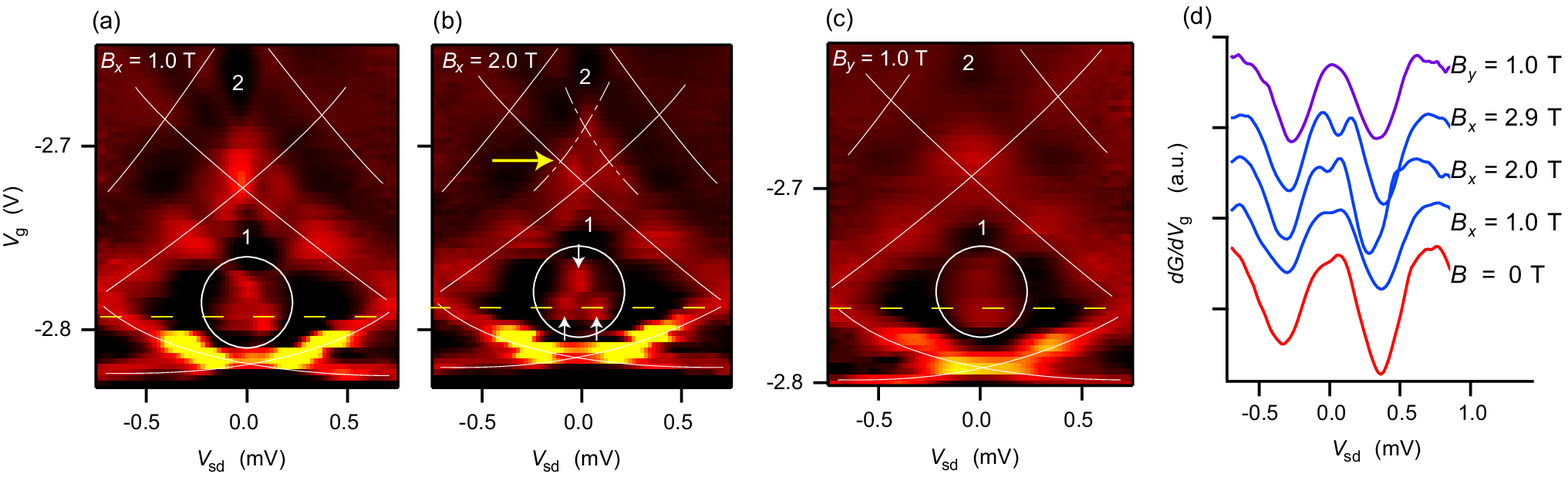}%
\caption{\label{fig_dGdVgBxCf}(Color online)\,Image plots of $dG/dV{\rm g}$ as a function of $V_{\rm sd}$ and $V_{\rm g}$ at (a) $B_x =1.0$, (b) $B_x=2.0$, and (c) $B_y=1.0$\,T.  Primary SBEs are indicated by the solid white lines. The three white arrows in (b) indicate three peaks inside the first diamond. In addition, the yellow arrow in (b) shows the Zeeman gap opening. (d)\,Line profiles of $dG/dV_{\rm g}$ at the lower peak in the first diamond (at the yellow broken lines in (a) through (c)) as a function of $V_{\rm sd}$ for $B =0, B_x=1.0, B_x=2.0, B_x=2.9$, and $B_y=1.0$\,T. For the $B_x = 2.9$\,T data, see Figure \ref{appendixfig_dGdVgBxCf} (c).  Each trace is offset for clarity.}
\end{figure*}

Subsequently, the energy spectroscopy for the channel under the double-layer QPC was measured.
Subband spacings of transverse modes at the QPC are observed in a spectroscopic measurement by controlling the Fermi energy $E_F$ through $V_{\rm g}$ and the chemical potentials between the source and drain $\Delta\mu_{\rm sd} = \mu_{\rm s} - \mu_{\rm d} = eV_{\rm sd}$.
Figure \ref{fig_dGdVgB0T} (a) shows the image plot of $dG/dV_{\rm g}$ as a function of $V_{\rm sd}$ and $V_{\rm g}$. 
The dark regions represent low $dG/dV_{\rm g}$; therefore, these regions indicate plateau regions in the conductance, whereas the brighter regions represent high $dG/dV_{\rm g}$, indicating that a Fermi energy passes through an SBE. 
It is to be noted that the pinch-off voltage is different from that in Fig.\,\ref{fig_G2K35mK} owing probably to unexpectedly localized electric charges. 
As compared to ordinary monolayer QPC cases\,\cite{Thomas_PRB1998,Kristensen,Cronenwett,Chen_NanoLett,Roessler}, or even several tunnel-coupled double-layer QPC cases\,\cite{Salis,Fischer,Smith}, the data reveal a rich SBE structure, particularly inside the first (lowest) SBE diamond (see also Fig.\,\ref{fig_dGdVgB0T} (b), which is an enlarged image plot of Fig. \ref{fig_dGdVgB0T} (a) around the first SBE structure).
In Figs.\,\ref{fig_dGdVgB0T}\,(a) and (b), we draw the SBE lines by connecting the maxima in $dG/dV_{\rm g}$ on the image plot with the primary integer series in solid lines.
The first large diamond appears from $V_{\rm g} \simeq -2.8$\,V and closes at $\simeq -2.7$\,V, with a width of approximately 1.5\,mV.
As is well known, this width is to determine the subband spacing in the QPC.
The electrostatic potential at the narrow constriction can be described as a saddle point model \cite{Glazman,Buttiker,LesovikSadovskyy} given by 
\begin{equation}
V(x,y)= V_0 -\frac{1}{2} m^\ast\omega_y^2y^2 + \frac{1}{2}m^\ast\omega_x^2x^2,   \label{eqVxy}
\end{equation}
where $V_0$ is the electrostatic potential at the saddle, and the confinement potential curvatures are expressed in terms of the harmonic oscillation frequencies $\omega_x$ and $\omega_y$.
It is to be noted that our coordinate is different with that used in Ref. \onlinecite{Buttiker}, in which the propagation direction is $x$.
The subband spacing in this diamond corresponds to $\hbar \omega_x=0.75$\,meV. % in that it matches to a rough estimation of 0.48\,meV from the harmonic oscillation potential $\hbar\omega_x$ at the QPC.
The observed diamond shapes resemble slightly crushed rhombuses as compared to those in previous reports\,(e.g., \cite{Kristensen}).
Subsequently, we focus on the small structures by drawing split SBE lines in the $dG/dV_{\rm g}$ result, using dash-dotted lines and a broken line.   
An enlarged image plot focusing on the structure in the first diamond is shown in Fig.\,\ref{fig_dGdVgB0T} (b).
From this experimental result, we observe three split SBE lines corresponding to the three peaks observed in Fig.\,\ref{fig_G2K35mK} (b) ($\alpha$, $\beta$ and $\gamma$) for the first-integer SBE.
We will demonstrate that this SBE splitting is supported by the in-plane magnetic fields dependence of $dG/dV_{\rm g}$.
Figure \ref{fig_dGdVgB0T} (c) shows the $G$ profiles in units of $G_0$ as a function of $V_{\rm sd}$. 
As shown, the conductance is asymmetric with respect to the positive and negative sides of $V_{\rm sd}$.
This asymmetry in $G$ is large below $G < G_0$. 
As an example of the asymmetric behavior, we show a line profile of $G$ at $V_{\rm g}=-2.766$\,V (the horizontal broken yellow line in Fig.\,\ref{fig_dGdVgB0T} (b)) with a red curve in Fig.\,\ref{fig_dGdVgB0T} (c).  
This asymmetric behavior was observed previously\,\cite{Kristensen}, and explained in terms of self-gating effects.
However, by analyzing the results of the shot noise measurements, which will be presented in Sec.\,\ref{ShotNoiseResult}, we inferred that this asymmetry has an intrinsic physical origin.

As we have explained in Sec.\,\ref{experiment}, the in-plane components of the magnetic field, $B_x$ and $B_y$, can be applied to the QPC independently. 
Figure \ref{fig_dGdVg_VgvsBxBy} shows the image plots of $dG/dV_{\rm g}$ as functions of (a) $V_{\rm g}$ and $B_x$, and (b) $V_{\rm g}$ and $B_y$.
As $B_x$ is increased with $B_y=0$\,T (fixed), each SBE except for the lowest SBE (marked with $\alpha$ in Fig.\,\ref{fig_dGdVg_VgvsBxBy} (a)) separates into two, then the upper branches move upwards. 
Even the SBE between the 0.5 and 1 plateaus decouples into two (marked with $\beta$ and $\gamma$).
Therefore, the SBE under the $G_0$ plateau splits into three, which is consistent with the observed SBE lines in Fig.\,\ref{fig_dGdVgB0T}.
The other SBEs show a Zeeman splitting similar to the cases of monolayer QPCs\,\cite{Graham,Hew_SpinIncoherent,Chen} as $B_x$ increases.
It is remarkable that the SBE splitting starts at approximately $B_x=1$\,T.
However, as shown in Fig.\,\ref{fig_dGdVg_VgvsBxBy} (b), the SBEs indicate no clear dependences on $B_y$ below 1\,T; instead, they decrease slightly, particularly for higher SBEs.
The lowest SBE shows no dependence of $B_x$ and $B_y$.
In addition, no clear onset of the second subladder (anti-symmetric wavefunction series) occurs for both in-plane fields below $G < 5G_0$, contrary to the previous double-layer QPC data\,\cite{Thomas_WaveFuncMixing,Salis,Fischer}.

Figures \ref{fig_dGdVgBxCf} (a) through (c) show the image plots of  $dG/dV_{\rm g}$ for $B_x = 1.0$, 2.0 ($B_y=0$\,T),
and $B_y=1.0$\,T ($B_x=0$\,T), respectively.
As $B_x$ increases, the structure in the first diamond (indicated by the white circles, SBE lines of $\beta$ and $\gamma$ in Figs.\,\ref{fig_dGdVgB0T} and \ref{fig_dGdVg_VgvsBxBy}) shows an interesting change.
The lower broad peak separates into two peaks gradually, in contrast to the upper peak that becomes a clear single peak.
This is demonstrated in Fig.\,\ref{fig_dGdVgBxCf} (b) ($B_x=2.0$\,T) as we indicate with three white arrows.
Meanwhile, at $B_y=1.0$\,T, each of the lower and upper peak smears out and becomes a broad peak.
In Fig.\,\ref{fig_dGdVgBxCf} (d), we plot the $dG/dV_{\rm g}$ profile of the lower peak at $V_{\rm g} = 2.795$\,V (indicated by yellow broken lines in Figs.\,\ref{fig_dGdVgBxCf} (a) to (c)) at $B=0, B_x=1, B_x=2.0, B_x=2.9$, and $B_y=1$\, T.
At $B=0$\,T, a small shoulder appears on the left side of the center peak (at $V_{\rm sd} = 0$\,mV).
However, we observe two peaks at $B_x=2.0$ and 2.9\,T clearly, and at $B_x=1.0$\,T slightly.
Thus, the observed structure inside the first diamond shows a clear dependence on the magnitude of $B_x$.
Meanwhile, the higher SBE in Fig.\,\ref{fig_dGdVgBxCf} (b) (indicated by the yellow horizontal arrow at $V_{\rm g}=-2.705$\,V) change differently; they exhibit a small diamond structure in accordance with the Zeeman gap opening as $B_x$ increases (see also Fig.\,\ref{appendixfig_dGdVgBxCf} in Appendix \ref{supplemental_dGdVg}).

%%% FIGURE 6 %%%
\begin{figure} [b]
\includegraphics[width=1\linewidth]{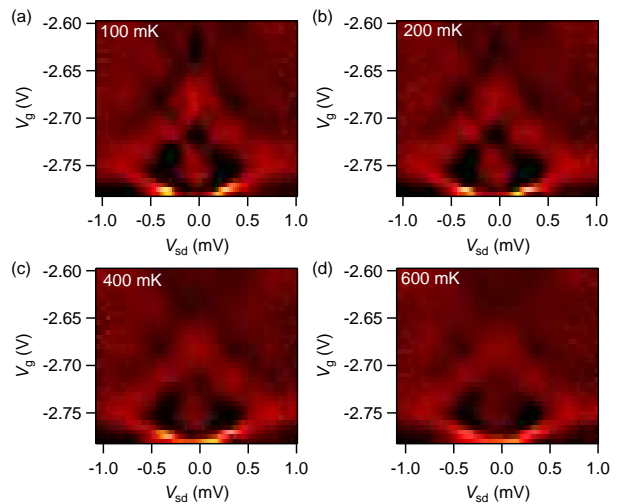}	
\caption{\label{fig_dGdVg_temp}(Color online)\,Image plot of $dG/dV_{\rm g}$ as a function of $V_{\rm sd}$ and $V_{\rm g}$ at $B=0$\,T for $T=$ (a) 100, (b) 200, (c) 400, and (d) 600\,mK. }
\end{figure}

In addition, we observe a result that is different from the previous results of the 0.7 anomaly. 
Figure \ref{fig_dGdVg_temp} shows the image plots of $dG/dV_{\rm g}$ for several temperatures from 100\,mK to 600\,mK.
Interestingly, the structure inside the first diamond smears out as $T$ is increased, showing a broad vague peak at the center of the diamond.
Therefore, it is clear that the structure observed in this study originates from the band-dispersion of the double-layer system.
Conversely, the $dG/dV_{\rm g}$ minimum for $0.5 G_0$ plateau is robust.
$G$ forms a clear plateau at $0.5 G_0$; after this plateau it increases without forming additional clear plateaus.

\subsection{\label{ShotNoiseResult}Results of the Shot Noise Measurement}

%%% FIGURE 7 %%%
\begin{figure} 
\includegraphics[width=1\linewidth]{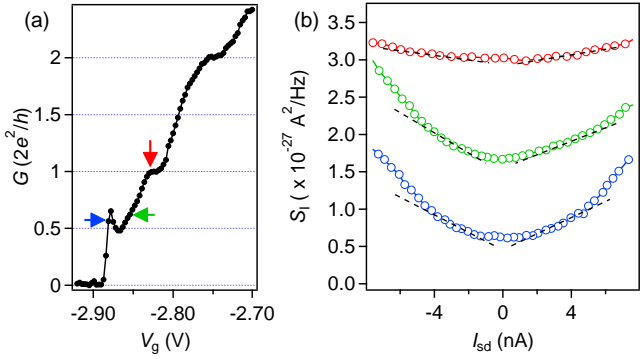}	
\caption{\label{fig_GandSI}(Color online)\,(a) $G$ as a function of $V_{\rm g}$ at $T=300$\,mK and $B=0$\,T for $V_{\rm sd}=0$\,V. (b)  $S_I$ as a function of $I_{\rm sd}$ for $V_{\rm g} = -2.88, -2.85$, and $-2.83$\,V (from the bottom trace to top). Each trace is offset for clarity. Colors of the traces correspond to the colors of the arrows in (a). }
\end{figure}

\begin{figure} 
\includegraphics[width=0.8\linewidth]{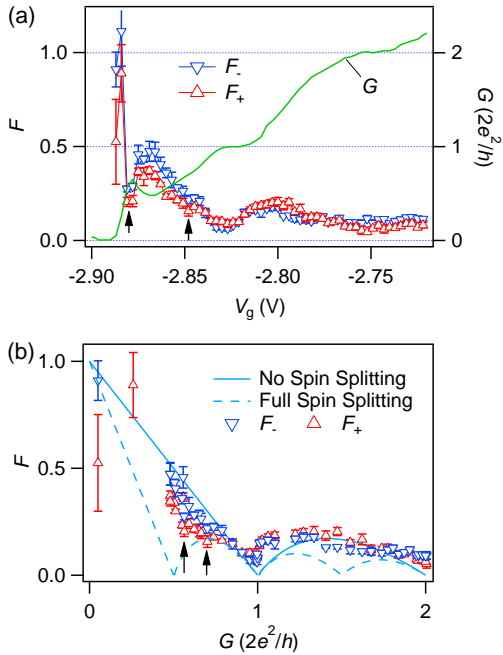}	
\caption{\label{fig_Fano0T}(Color online)\,(a) $F_+$ and $F_-$ (left axis) and $G$ (right axis) as a function of $V_{\rm g}$ at $B=0$\,T. (b) $F_+$ and $F_-$ as a function of $G$. The solid  lines and broken lines represent the theoretically expected values of the Fano factor for no spin splitting and full spin splitting, respectively. Fano factor reductions at $G=0.5 G_0$ and $0.7G_0$ are indicated by the upper arrows.}
\end{figure}

To further obtain information on the phenomenon from a different aspect, we performed shot noise measurements.
Figure \ref{fig_GandSI} (a) shows $G$ as a function of $V_{\rm g}$ at 300\,mK. 
The overshoot observed at the $0.5G_0$ plateau is more prominent at higher temperatures, resembling the one observed in \cite{Sfigakis}.  
We attribute the appearance of this overshoot to a resonance mode due to the superimposed transmission and reflection on the lowest SBE at the QPC region.
Figure \ref{fig_GandSI} (b) shows $S_{\rm I}$ as a function of $I_{\rm sd}$ for $V_{\rm g}=-2.88, -2.85$, and $-2.83$\,V.
$S_{\rm I}$ shows a parabolic behavior for $|eV_{\rm sd} | \lesssim 2k_{\rm B} T$, and then shows a linear dependence for $|eV_{\rm sd} | \gtrsim 2k_{\rm B} T$, which is a typical behavior of the shot noise with crossover from thermal noise to shot noise.
We observe an asymmetric dependence between the positive and negative $I_{\rm sd}$ near $G=0.7G_0$, which was also observed previously\,\cite{Roche,Kristensen} and explained in terms of the self-gating effect in QPC. 
However, this asymmetry in $S_{\rm I}$ is observed only for $ -2.875 \leq V_{\rm g} \leq -2.84$\,V (for $0.5 G_0 < G < G_0$), and does not occur in other $V_{\rm g}$ values, thus suggesting other possibilities.
Accordingly, the slope of $S_{\rm I}$ is always higher in the negative side of $I_{\rm sd}$ for $0.5 G_0 < G < G_0$.
As we have stated earlier, we derived the Fano factor from the slope of $S_{\rm I}$ as $F = S_{\rm I}/(2e\langle I_{\rm sd} \rangle)$.
Owing to the asymmetry between the positive and negative $I_{\rm sd}$ sides of the $S_{\rm I}$,
we used the Fano factor of the positive side $F_+$ and negative side $F_-$, and plotted them as a function of $V_{\rm g}$, as shown in Fig.\,\ref{fig_Fano0T} (a).
Further, the zero bias ($V_{\rm sd} =0$) conductance $G$ is plotted on the right axis in Fig.\,\ref{fig_Fano0T} (a).
Consistent with the $S_{\rm I}$ result, $F_-$ is larger than $F_+$ between the $0.5 G_0$ and $ G_0$ plateaus.

%%%% Figure 8 %%%%%
%

%%% FIGURE 9 %%%
\begin{figure} 
\includegraphics[width=0.8\linewidth]{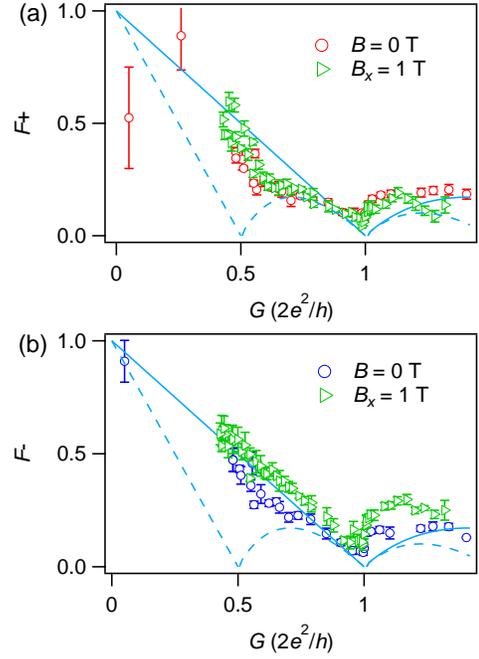}	
\caption{\label{fig_Fano_CfBx}(Color online)\,(a) $F_+$  and (b) $F_-$ as a function of $G$ for $B=0$\,T and $B_x=1$\,T. The solid lines and broken lines represent the same as those in Fig.\,\ref{fig_Fano0T} (b).}
\end{figure}

In a noninteracting scattering process, theory predicts\cite{Blanter}
\begin{equation}
F = \frac{\sum_{n}T_n (1-T_n)}{\sum_n T_n},  \label{EqFano_theory}
\end{equation} 
where $T_n$ denotes the transmission probability of the $n$-th channel. 
We replot $F_+$ and $F_-$ as a function of $G$ in Fig.\,\ref{fig_Fano0T} (b), along with the theoretical value of $F$ when no spin splitting (the solid lines) and full spin splitting (the broken lines) occur. 
Both $F_+$ and $F_-$ are suppressed at $G = G_0 $ and $2 G_0$, thus implying the formation of a single perfect conductance channel in the coupled DQW for the plateau region.
Two important features of $F_+$ and $F_-$ observed are 1) a clear suppression at $G = 0.5  G_0$ and a rapid increase after this reduction as $G$ is decreased, and 2) a small reduction at $G \sim 0.7 G_0$ (both reductions are indicated by the upper arrows in Fig.\,\ref{fig_Fano0T} (a) and (b)).
Regarding the first point, the decrease in the Fano factor indicates that $E_{\rm F}$ finishes crossing an SBE. After the suppression at $0.5G_0$, the Fano factor is increased even when the plateau of $G$ is established.
Generally, the increase in the Fano factor indicates that a new conduction channel opens as $G$ increases from $G= 0.5 G_0$.
The second point suggests that, as shown previously\,\cite{Roche,DiCarlo,Nakamura} regarding the 0.7 anomaly,  the existing channels' transmission probabilities contribute unequally to the conductance. 
This small reduction appears for both $F_+$ and $F_-$.
The $F$ values are larger than the theoretical values of $F$ at the conductance plateau region.
For the enhanced Fano factor, three possibilities can be considered:
electron heating\,\cite{Muro}, channel mixing, and $1/f$ noise.
However, the $1/f$ noise scarcely contribute to the enhancement in this experiment owing to the noise measurement technique using a high resonant frequency $LC$ circuit and double-high electron mobility transistor amplifier\,\cite{Arakawa_APL}.

Furthermore, we measured the shot noise in the presence of in-plane magnetic fields.
Fig.\,\ref{fig_Fano_CfBx} shows $F_+$ and $F_-$ against $G$ for $B=0$ and $B_x = 1$\,T.
In the presence of in-plane magnetic fields, the Fano factor increases.
At $B_x=1$\,T, the difference between $F_+$ and $F_-$ becomes larger than the zero field difference between the 0.5 and 1 plateau regions. 
As a notable difference, $F_-$ obeys the theoretical dependence well.
%We will revisit this $B_x$ dependence in Sec.\,\ref{SOISplitting}.

\section{\label{discussion}Discussion}

In this section, first, we summarize our observations before presenting a discussion of  the results.
First, it is shown that three maxima exist inside the first diamond for the $dG/dV_{\rm g}$ result, especially in the presence of a large $B_x$.
Next, $G$, $dG/dV_{\rm g}$ and $F$ exhibit an asymmetric dependence with respect to $V_{\rm sd}$.
However, in our results, an apparent beginning of the second layer SBE such as those observed in Refs.\,\cite{Thomas_WaveFuncMixing,Salis,Fischer} is not observed contrary to expectation.
We cannot completely deny the possible effects from double-layer wavefunction mixing on the issues above.
Thus, we must specify whether our observation originated from double-layer wavefunctions.
Hence, we conducted computer simulations using the nextnano simulation software\,\cite{nextnano}. The simulation results do not support the formation of double-layer wavefunctions; thus, it is difficult to explain the results solely based on double-layer effects. 
Having obtained the simulation results, we propose a possible explanation for the experimental results above using the spin effect, i.e., the SOI-modified dispersion relation in particular.

\subsection{\label{Simulation}Simulation Results}

%%% FIGURE 10 %%%
\begin{figure} 
\includegraphics[width=0.8\linewidth]{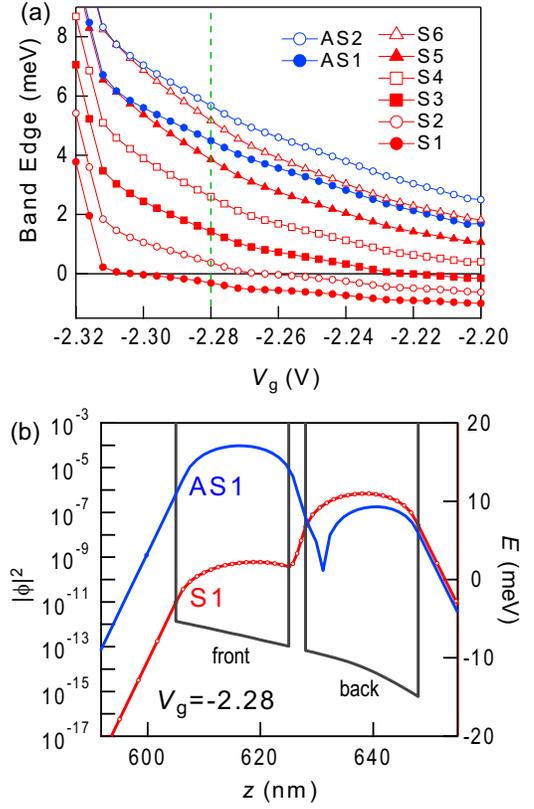}	
\caption{\label{fig_SBE_WaveFunc}(Color online)\,(a) $V_{\rm g}$ dependence of the SBEs. S and AS denote symmetric and anti-symmetric wavefunctions, respectively, and the number index represents the $m$-th lateral mode. (b) $|\phi|^2$ for S1 and AS1 at $V_{\rm g} = -2.28$\,V (the dotted green line on the simulation result of (a)). The black line represents the quantum well potential $V(z)$ for this gate voltage value. The origin of $z$-axis starts from the sample surface.  }
\end{figure}
 
Because the system contains two layers (front and back), we must consider two subladders for the wavefunctions and confinement potentials.
We denote the wavefunction of the system as 
\begin{equation}
\Psi_{l,m}(x,y,z) = u(y)\psi_{l,m}(x,z)     \label{WavefunctionPsi}
\end{equation}
with direction $y$ for propagating modes, and directions $x$ and $z$ for lateral and vertical (quantum well) confinement, respectively.
The envelope wavefunction can further be denoted as  
\begin{equation}
\psi_{l,m}(x,z) = \chi_m(x)\phi_l^p(z),
\end{equation}
where $\chi_m(x)$ denotes the $m$\,th lateral mode and $\phi_l^p(z)$ denotes the $l$\,th vertical wavefunction in the quantum well.
For tunnel-coupled vertical modes, 
\begin{equation}
\phi_l^p(z) = \alpha \varphi_l^{\rm f}(z) + \beta e^{i\theta}\varphi_l^{\rm b}(z),\ \ \ \  \alpha^2+\beta^2=1
\end{equation}
where $\varphi^{\rm f}$ and $\varphi^{\rm b}$ denote the wavefunction in the front and back layers (subladder index), respectively, and $\theta$ denotes the interlayer phase difference.
The index $p$ uses S or AS : for $p= {\rm S}$, $\theta=0$ for the symmetric bonding state, and for  $p= {\rm AS}$, $\theta =\pi$ for the anti-symmetric bonding state.

To confirm the SBEs in the first diamond, the wavefunction energies at the QPC were simulated using the self-consistent Schr\"{o}dinger-Poisson method with nextnano.
We first performed a one-dimensional (1D) simulation in the $z$ direction with reference to the characteristics of the bulk, i.e., the calculated $\Delta_{\rm SAS}$ and $V_{\rm g}$ dependence of $G$ to determine the simulation parameters (see Appendix \ref{SupplementalSimulation}).
Subsequently, we proceeded with two-dimensional (2D) simulations in the $xz$ plane as a function of $V_{\rm g}$.
The SBE energies are calculated as the eigenvalues of the quantized wavefunctions in the $xz$ plane under a lateral parabolic confinement potential.
It is noteworthy that although the 2D simulation did not consider the $y$ direction, we assume that the $y$-directional eigen-energies exhibit a qualitatively equivalent dependence on $V_{\rm g}$ in the QPC region.  
Thus, the lateral potential and width are determined based on the $V_{\rm g}$ value.
Figure \ref{fig_SBE_WaveFunc} (a) shows the SBE energies as a function of $V_{\rm g}$ at the center of the QPC region. 
We found that the energy of the lowest anti-symmetric wavefunction ($l=1, p={\rm AS}, m=1$) was higher than that of the fifth symmetric wavefunction ($l=1,p={\rm S}, m=5$), because the screening effect of the front layer was extremely strong to allow for the electrons to realize the anti-symmetric wavefunction (hereinafter, we denote wavefunction using two indexes, $p$ and $m$, such as AS1, because $l$ is always 1). 
In Fig.\,\ref{fig_SBE_WaveFunc} (b), we show the $|\phi|^2$ of the lowest symmetric wavefunction (S1) and the anti-symmetric wavefunction (AS1) at the first plateau region.
The wavefunction shows a large imbalance between the front and back layers, indicating an extremely weak coupling between the two layers under an applied strong electric field of approximately $\sim 4$\,V/($\mu$ m).
Hence, we expect electrons to exist primarily in the back layer and their wavefunction to permeate to the front layer; thus, the system behaves as a single-layer system with a large potential gradient toward the front layer.

%%%%%%%%%%%%%%%%%%%%%%%%%%%%%%%%%

%%% FIGURE 11 %%%
\begin{figure*} 
\includegraphics[width=0.66\linewidth]{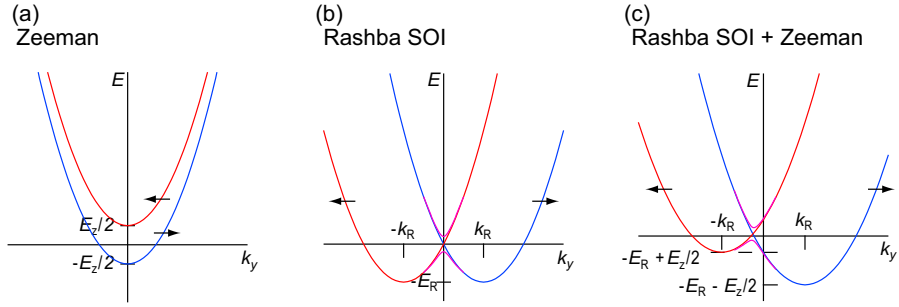}	
\caption{\label{Fig_ZeemanRashba}(Color online)\,Dispersion relations for (a) Zeeman splitting, (b) Rashba SOI splitting, and (c) Rashba SOI plus Zeeman splitting. }
\end{figure*}

\subsection{\label{SOISplitting} Possible Explanation with SOI-induced Split Dispersion Relation}

To explain the structure in the first diamond (indicated by the white circles in Fig.\,\ref{fig_dGdVgBxCf}), the following simple relationship between the density of states (DOS) and conductivity can be useful.
As is well known, the ballistic electron transport in a QPC shows the conductance that changes stepwise depending on the number of subbands below the Fermi level.
Each subband carries the current 
\begin{equation}
j = e^2V_{\rm sd}n(E)v(E)        \label{j}
\end{equation}
where $n(E) = \frac{1}{2\pi}\frac{\partial k}{\partial E}$ denotes the 1D unidirectional density of states, and $v=\frac{2\pi}{h}\frac{\partial E}{\partial k}$ denotes the group velocity.
Therefore, cancellation between the DOS and the Fermi velocity causes the conductance quantization.
Equation (\ref{j}) describes the importance of the DOS, because the conductance is the result of the integral of the current divided by the applied voltage.
Experimentally, a sudden DOS change results in a large conductance jump and a large transconductance peak.
In our experiment, the brighter the SBE in the $dG/dV_{\rm g}$ plot, the larger are the DOS changes. 
Therefore, we observed three large DOS changes within the first diamond, as shown explicitly in Fig.\,\ref{fig_dGdVgBxCf} (b).

For the candidate of the threefold DOS change, we suggest the dispersion relation that splits in the wavenumber $k$ direction, such as the SOI-induced splitting\,\cite{Goulko,Quay,Kammhuber} and the in-plane magnetic-field-induced splitting for tunnel-coupled double-layer systems\,\cite{Lyo_1994}, because three minima appear in the subbands.
However, taking into account the simulation result, the possibility of realizing an in-plane magnetic-field-induced splitting is highly unlikely, because well-developed tunnel-coupled wavefunctions are a prerequisite for this to occur (we will discuss this in detail later). 
Regarding the SOI in this case, the space inversion symmetry is expected to be maintained for the $x$ and $y$ directions, but broken for the $z$ direction. 
Thus, the Rashba SOI\,\cite{Rashba} with regard to the potential gradient in the $z$ direction and the current in the $y$ direction ($[0, 0, \partial V(z)/\partial z] \times [0, k_y, 0] \parallel B_x$) is expected.
The Hamiltonian regarding the Rashba SOI with this broken symmetry is
\begin{align}
\mathcal{H} &= \frac{\hbar^2k_y^2}{2m^\ast} -\frac{\hbar^2}{4{m^\ast}^2c^2}\sigma_x \frac{\partial V(z)}{\partial z}k_y  \\
&= \frac{\hbar^2k_y^2}{2m^\ast} + \alpha_{\rm R}\sigma_xk_y,      \label{Hamiltonian}
\end{align}
where $V(z)$ denotes the potential function of the DQW, $\sigma_x$ denotes the $x$ component of the Pauli matrix, and $\alpha_{\rm R}$ is the so-called Rashba parameter. 
From Eq.\,(\ref{Hamiltonian}) above, we can derive the dispersion relation with the Rashba SOI as 
\begin{equation}
E^\leftrightarrows(k_y) = \frac{\hbar^2k_y^2}{2m^\ast} \pm \alpha_{\rm R}k_y.  \label{dispersion}
\end{equation}
Then, the energy assumes a minimum value of $ - \hbar^2k_{\rm R}^2/(2m^\ast) = -E_{\rm R}$ at $k_y= \mp \frac{m^\ast \alpha_{\rm R}}{\hbar^2} = \mp k_{\rm R}$.
Further, according to analysis\,\cite{Goulko,Quay}, the $k$-directional split subbands are mixed; consequently, the subbands repel and open a gap into the upper and lower branches (see Fig.\,\ref{Fig_ZeemanRashba} (b)).
Importantly, the lower branch contains two minima and the upper branch contains one minimum, at which the up- and down-spin DOSs are degenerated; hence, this SOI-modified dispersion exhibits three large DOS changes.
Furthermore, in the presence of $B_x$, Eq.\,\ref{dispersion} is modified as follows:
\begin{equation}
E^\leftrightarrows(k_y) = \frac{\hbar^2k_y^2}{2m^\ast} \pm \alpha^\prime_{\rm R}k_y \pm \frac{1}{2}g^\ast\mu_{\rm B}B_x,  \label{dispersion_RashbaZeeman}
\end{equation}
where $\mu_{\rm B}$ denotes the Bohr magneton.
The dispersion relations of the Zeeman splitting, Rashba SOI splitting, and Rashba SOI plus Zeeman splitting cases are illustrated in Fig.\,\ref{Fig_ZeemanRashba}.
The Rashba parameter should be modified because of an additional magnetic confinement potential  created by $B_x$, $m^\ast\omega_{B_x}^2z^2/2$\,\cite{Salis} ($\omega_{B_x} =eB_x/m^\ast$) in the $yz$ plane, as follows:
\begin{equation}
\alpha^\prime_{\rm R} = \frac{\hbar^2}{4{m^\ast}^2c^2}\frac{\partial}{\partial z} \left[ V(z) +\frac{1}{2}m^\ast\omega_{B_x}^2z^2 \right].
\end{equation}
Thus, the Rashba energy increases with the increase in $B_x$, which is a magnetic field parallel to the Rashba SOI field.
This indicates that the two minima in the dispersion curves of Rashba SOI separate with the increase in $B_x$; 
further, the crossing point and a side of a minimum separate vertically, whereas the other side approaches.
As shown in Fig.\,\ref{fig_dGdVgBxCf}, the lower two maxima inside the first diamond separate as $B_x$ increases, and thus agree qualitatively to the behavior of minima in the dispersion curves of Rashba SOI.

%%% FIGURE 12 %%%
\begin{figure} [b]
\includegraphics[width=1\linewidth]{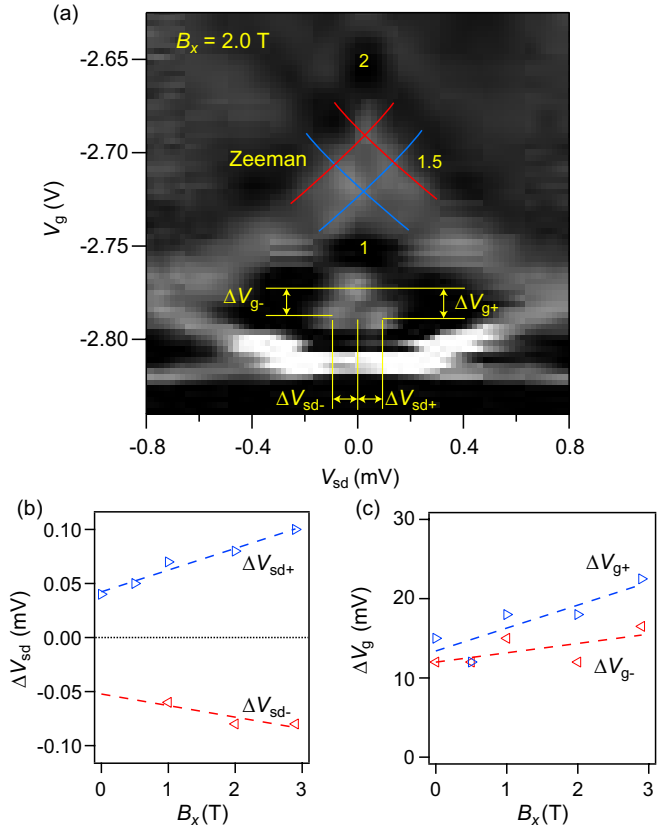}	
\caption{\label{dGdVg2T_and_VsdVgvsBx}(Color online)\,(a) Enlarged image plot of $dG/dV_{\rm g}$ at $B_x=2.0$\,T. Two sets of Zeeman splitting SBE lines have been indicated. (b) Plots of $\Delta V_{\rm sd}$ and (c) $\Delta V_{\rm g}$ as a function of $B_x$.}
\end{figure}

We extract the positions of the lower two maxima as $\Delta V_{\rm sd +}$ and $\Delta V_{\rm sd -}$.
In addition, the separation of the center maximum and each lower maximum is extracted as $\Delta V_{\rm g +}$ and $\Delta V_{\rm g -}$ (see Fig.\,\ref{dGdVg2T_and_VsdVgvsBx} (a) for graphical illustration). 
Figure \ref{dGdVg2T_and_VsdVgvsBx} (b) and (c) show the $\Delta V_{\rm sd}$ and  the $\Delta V_{\rm g}$ values, respectively, as a function of $B_x$.
Although $\Delta V_{\rm g -}$ increases slightly, the overall changes correspond well to the three points in the dispersion curves of the Rashba SOI plus Zeeman splitting---the crossing point and the two minima.
Therefore, the three maxima observed inside the first diamond can be attributed to these points.  
Considering that the Rashba SOI field is proportional to $\frac{\partial V(z)}{\partial z}p_y$, the principle behind the observed SOI is simple: the strong potential gradient and high mobility (or the large relaxation time\,\cite{Edelstein}) of the sample.
In our opinion, the center barrier in the DQW produces this strong potential gradient, as shown in the potential profile $V(z)$ in Fig.\,\ref{fig_SBE_WaveFunc} (b).

Furthermore, the shot noise results support the conjecture above in that the SBE splitting originates from the SOI. 
As shown in Fig.\,\ref{fig_Fano_CfBx}, the additional $B_x$ increases $F_-$ to theoretical values.
In addition, the difference between $F_-$ and $F_+$ becomes larger at $B_x=1$\,T.
Given that $B_x$ is in the same direction as that of the effective Rashba magnetic field $B_{\rm eff}$, when the current flows from the source to drain, $V_{\rm sd} >0$ (hence the electron momentum is in the opposite direction), a positive $B_x$ supports $B_{\rm eff}$. 
However, the situation is completely different when $V_{\rm sd}$ is negative, because a positive $B_x$ cancels $B_{\rm eff}$ as $B_{\rm eff}$ is induced to the negative $x$ direction.
Therefore, in the presence of the positive $B_x$, the separation by the Rashba SOI is enhanced for $V_{\rm sd} >0$ and decreased for $V_{\rm sd} < 0$.
Consequently, $G$ is suppressed for $V_{\rm sd}>0$ and hence $F_+$, and vice versa.
As shown in Fig.\,\ref{fig_Fano0T}, this anisotropic Fano factor is observed at 0\,T. 
This is attributed to the effective Zeeman energy $g\mu_{\rm B}B_{\rm eff}$.

An alternative SOI-like dispersion splitting can be considered in a tunnel-coupled double-layer system.
According to Ref\,\cite{Lyo_1994}, an in-plane field induces the subband splitting in proportional to the magnitude of the in-plane field in the direction perpendicular to the in-plane field for 2DEG systems.
Thus, $B_x$ splits the subband in the $k_y$ direction as $\Delta k_y = d/(\hbar/(eB_x))$. 
However, the estimated separation for $B_x =1$\,T is $\Delta k_y = 3.5 \times 10^7$\,m$^{-1}$, thus yielding $\frac{(\hbar\Delta k_y)^2}{2m^\ast} = 0.69$\,meV.
Although the theory considers a double-layer 2DEG system, this value is significantly large, comparable to the observed first diamond splitting.
Furthermore, we cannot explain the small split that is already observed at the zero magnetic field.
In addition, a strong double-layer coupling is a prerequisite for this splitting.
As shown in Fig.\,\ref{fig_SBE_WaveFunc} (b), the wavefunctions in the lower subbands are the highly unbalanced bonding state.
Therefore, this cannot be the primary contribution to the horizontal splitting.

Finally, we would like to briefly discuss the reentrant conductance behavior that was observed in strong SOI systems in previous studies\,\cite{Quay,Kammhuber,Heedt}.
In this study, a small reentrant feature was confirmed as shown in Fig.\,\ref{fig_G2K35mK} (b), and in the conductance data in Fig.\,\ref{fig_GandSI} and \ref{fig_Fano0T}, although we interpreted them as a resonant mode.
However, these features are not apparent compared with those in Ref.\cite{Quay,Kammhuber,Heedt}.
We attribute this to the band structure of the sample: the second lowest band exists immediately above the lowest band. This configuration suppresses ``the helical gap'' and obscures the reentrant behavior. 
%We will observe a reentrant conductance in a fully lifted-degeneracy situation\,\cite{bilayerQPCBz}.

\section{Concluding Remarks and Perspectives}
\label{conclusion}

We herein have revealed the SBE lines as a consequence of the wavenumber direction subband splitting induced by a strong SOI.
We have observed the coexistence of a 0.5$G_0$ plateau and a structure at 0.7$ G_0$ in a double-layer QPC system.
The structure observed in the $dG/dV_{\rm g}$ spectroscopy has revealed three maxima corresponding to the three minima in the dispersion relation of the wavenumber-directional subband splitting.
We attribute this splitting to a strong SOI owing to the high potential gradient at the center barrier and the high mobility of the double-layer sample.
The Fano factor obtained from the shot noise measurement have indicated an asymmetric transmission probability.
This result further supports the SOI-modified dispersion model and the asymmetry observed in the conductance measurement.
%Moreover, we have proposed an equation that relates the wavenumber propagating through the QPC (in our case $k_y$) and $V_{\rm sd}$.
%This further deduces the SBE lines in a $V_{\rm sd} - V_{\rm g}$ plane; as a consequence, we have proposed a possible explanation for the `0.7 anomaly' as well.
However, multiple unanswered questions still exist that require theoretical considerations and additional experiments.
This experiment includes useful information on spintronics and quantum engineering that would benefit applications.
In particular, a strong SOI in a GaAs/AlGaAs sample invokes spintronic applications in this well-developed platform.
In addition, we intend to perform shot noise measurements in the QHE region of this system in the future.

\appendix
\label{appendix}

\section{Shubnikov de-Haas Oscillation Analysis} 
\label{SdH}

%%% FIGURE 14 SdH %%%
\begin{figure} [b]
\includegraphics[width=1\linewidth]{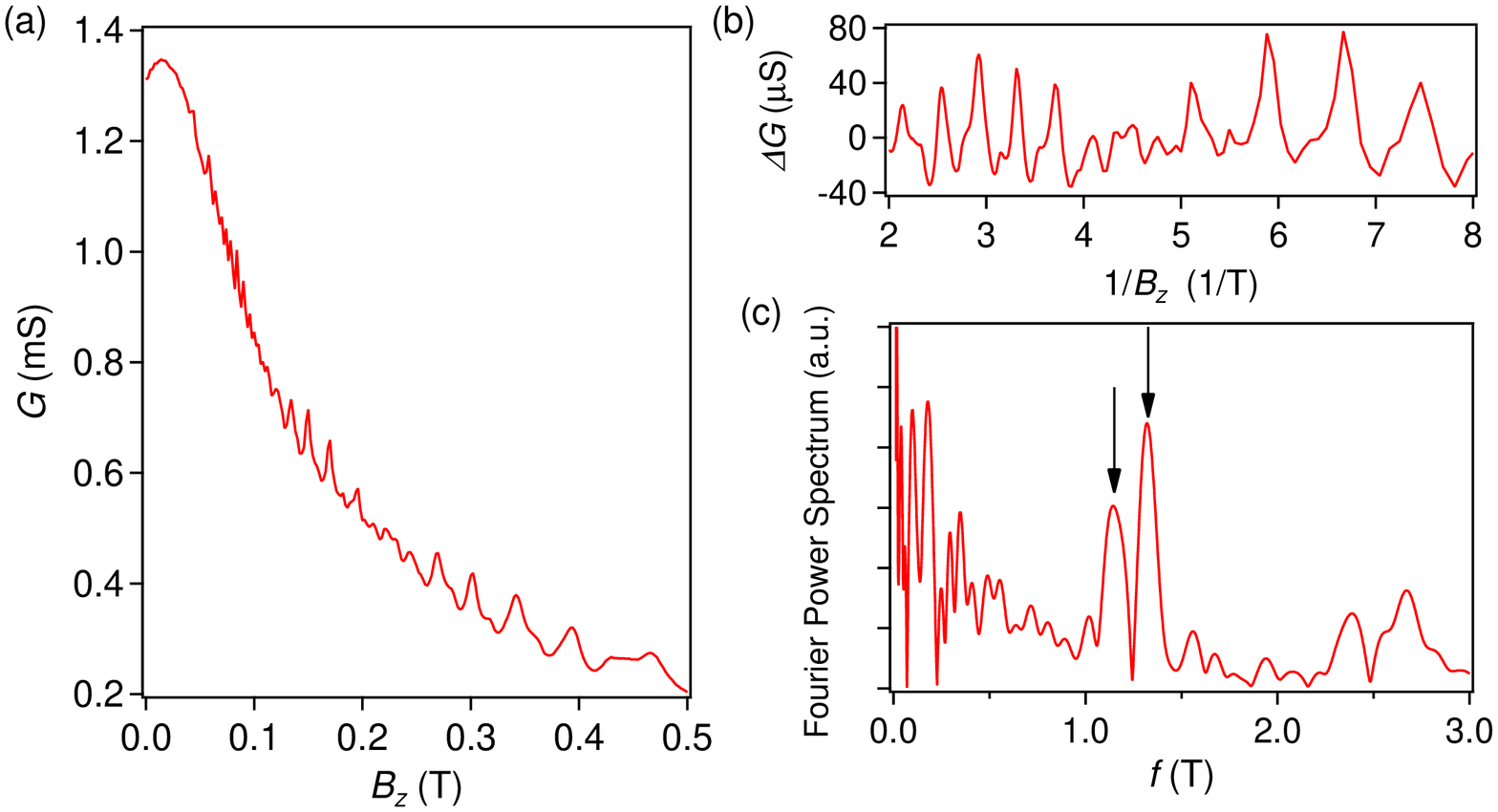}
\caption{\label{fig_SdH}(Color online)\,(a) $G$ as a function of $B_z$. (b) SdH oscillation extracted from (a). $\Delta G$ represents the conductance subtracted the back ground conductance change. (c) FFT power spectrum of the data in (b). }
\end{figure}

First, we measure the Shubnikov de-Haas (SdH) oscillation at zero bias ($V_{\rm sd}=0$) and zero split gate voltages ($V_{\rm g}=0$) in low magnetic fields at the lowest temperature available in this experiment, to obtain the electron densities and tunnel coupling strength between the layers.
Figure \ref{fig_SdH} (a) shows $G$ as a function of $B_z$. 
As a clear sign of the weak localization effect\,\cite{Bergmann}, positive magneto-conductance is observed initially.
Subsequently, the difference in density between the symmetric state and anti-symmetric state results in a beating of the SdH oscillations in $G$\,\cite{Boebinger}.
This beating is resolved into two sharp peaks of Fourier power spectrum from the fast Fourier transform analysis of the $1/B_z$ dependence of $G$, as shown in Fig.\,\ref{fig_SdH} (c) by the arrows.
The density $\rho$ corresponding to each peak is, as we mentioned earlier, $0.64 \times 10^{11}$ and $0.56 \times 10^{11}$ cm$^{-2}$ from a well-known relation between the SdH frequency $f=\Delta B_z$ and $\rho$, $\rho = 2ef/h$, and the energy separation between the symmetric and anti-symmetric states $\Delta_{\rm SAS}$ is $\Delta_{\rm SAS} = \pi \hbar^2(\rho_{\rm S} -\rho_{\rm AS})/m^\ast = 0.29$\,meV, where $m^\ast=0.067 m_e$ in GaAs with $m_e$ denotes the electron rest mass, and $\rho_{\rm S}$ and $\rho_{\rm AS}$ denotes the electron density in the symmetric and anti-symmetric states, respectively.

\section{Supplemental $dG/dV_{\rm g}$ Data}
\label{supplemental_dGdVg}

%%% FIGURE 15 %%%
\begin{figure*} 
\includegraphics[width=1\linewidth]{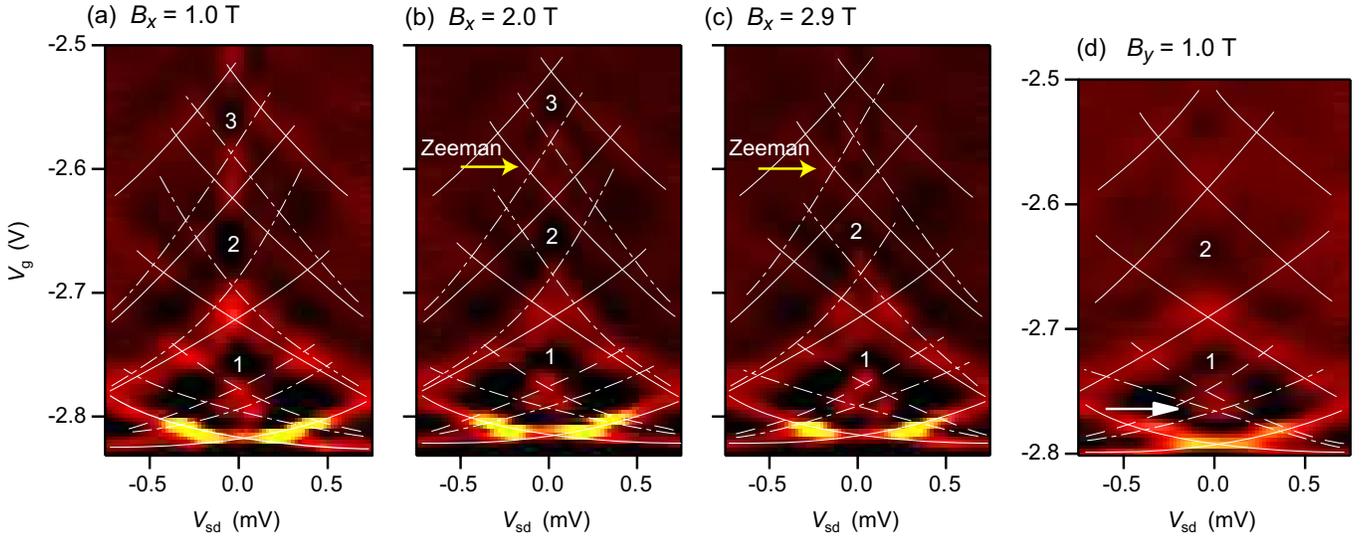}%
\caption{\label{appendixfig_dGdVgBxCf}(Color online)\,Overall view of image plots of $dG/dV{\rm g}$ as a function of $V_{\rm sd}$ and $V_{\rm g}$ at (a) $B_x =1.0$, (b) $B_x=2.0$, (c) $B_x=2.9$, and (d) $B_y=1.0$\,T. The yellow arrows in (b) and (c) show subband openings due to the Zeeman splitting. (e)\,Line profiles of $dG/dV_{\rm g}$ at the white arrows in (a) through (d) as a function of $V_{\rm sd}$ for $B =0, B_x=1.0, B_x=2.9$, and $B_y=1.0$\,T. Each trace is offset for clarity.}
\end{figure*}	

Figures \ref{appendixfig_dGdVgBxCf} (a) through (d) show the overall view of the image plots of $dG/dV_{\rm g}$ as a function of $V_{\rm sd}$ and $V_{\rm g}$ for $B_x = 1.0,2.0,2.9$\,T and $B_y=1.0$\,T, respectively.
For $B_x =2.0$ and 2.9\,T, the spin degeneracy is resolved for higher SBEs; consequently, we observe a minimum (dark region) corresponding to the $2.5G_0$ plateau (indicated by yellow arrows).
From this gap opening, the Zeeman splitting is $\approx 0.09$\,meV at $B_x=2.0$\,T. 
Compared to the bare $g$-factor of GaAs ($|g| = 0.44$), the Zeeman energy, $|g|\mu_{\rm B} B$, at this in-plane magnetic field is approximately twice that of the bare Zeeman splitting.

\section{Computer simulation using nextnano software}
\label{SupplementalSimulation}

%%% FIGURE 16 %%%
\begin{figure} 
\includegraphics[width=1\linewidth]{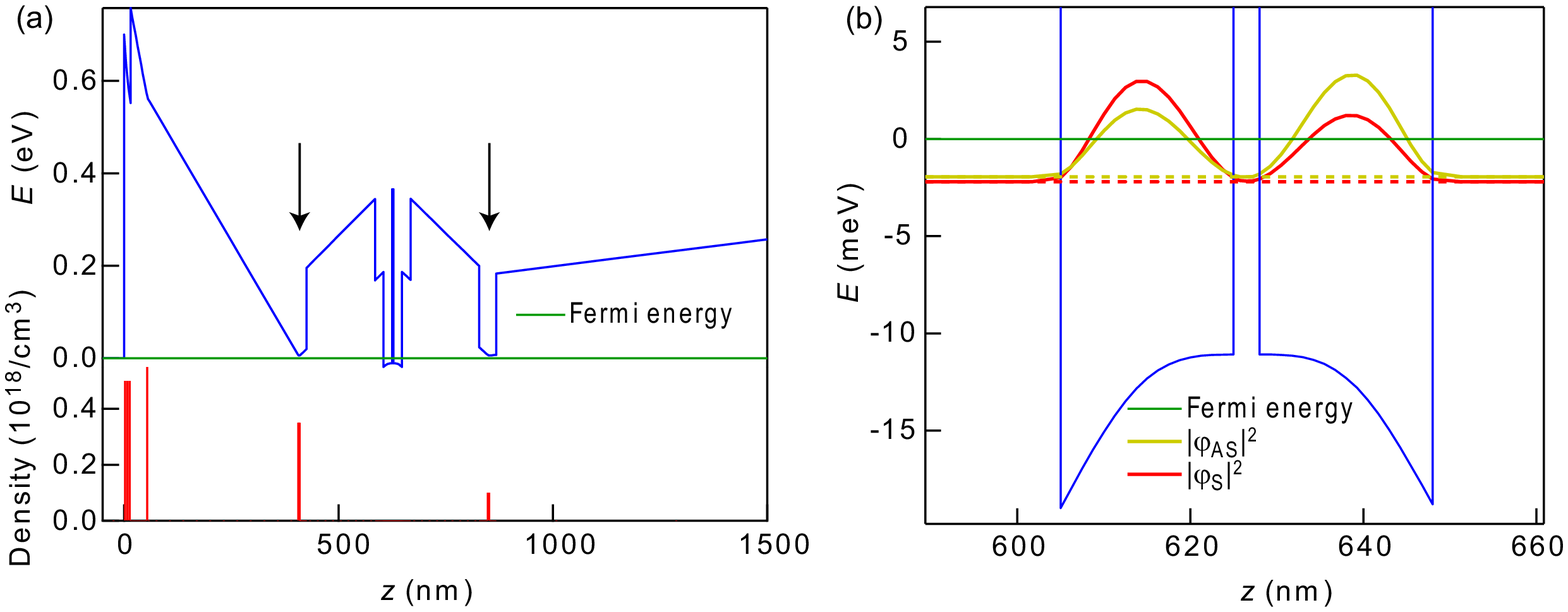}	
\caption{\label{fig_E}(Color online)\,(a) Potential distribution for the $z$ direction of the sample (upper) and the electron density profile (lower). The two downward arrows indicate the positions of $\delta$-doping. (b) Probability densities for the lowest two energy wavefunctions (symmetric $\varphi_{\rm S}$ and anti-symmetric $\varphi_{\rm AS}$ state) at the DQW confinement for the $z$ direction. }
\end{figure}

%%% FIGURE 17 %%%
\begin{figure}
\includegraphics[width=0.8\linewidth]{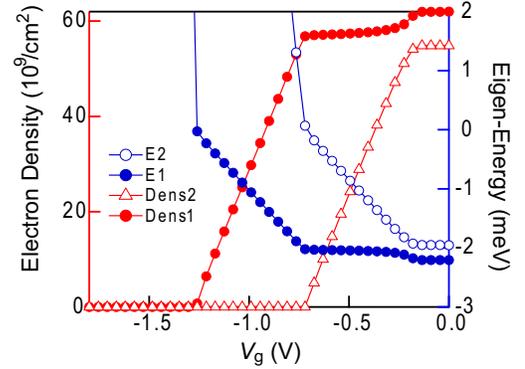}	
\caption{\label{fig_ElectronDensityAndEigenEnergy}(Color online)\,1D eigen-energies and electron densities for each layer as a function of $V_{\rm g}$. E and Dens represent eigen-energies and electron densities, respectively; 1 and 2 correspond to the back layer and front layer, respectively.}
\end{figure}

To estimate the double-layer effects on the conductance, we must calculate the wavefunctions at the double-layer QPC under a strong electric field confinement.
Hence, we used the electronic simulator software, nextnano\,\cite{nextnano}.
To supplement the main text, we provide the 1D simulation results of the $\Delta_{\rm SAS}$ calculation and the $V_{\rm g}$ dependence of the wavefunctions. 
Figure \ref{fig_E} (a) shows the potential profile for the $z$ direction and the electron density profile.
Owing to our careful design, two Si $\delta$-doping positions, indicated by the two downward arrows, render the DQW symmetric against the $z$ direction successfully. 
Fig.\,\ref{fig_E} (b) shows the energy of symmetric and anti-symmetric wavefunctions and their probability density profiles at $V_{\rm g}=0$\,V.
The tunnel gap, $\Delta_{\rm SAS}$, is calculated as 0.25\,meV, which is extremely close to the experimental value.
We tabulate the measured and calculated values of $\Delta_{\rm SAS}$ in Table \ref{Delta_SAS}, along with the densities of the lowest symmetric and anti-symmetric wavefunctions.

Figure\,\ref{fig_ElectronDensityAndEigenEnergy} shows the calculated eigen-energies from the 1D simulation ($z$-direction) for the lowest wavefunction and the electron density for each layer as a function of $V_{\rm g}$.
As $V_{\rm g}$ increases in the negative direction, the potential of the front layer increases, the symmetric state electrons depopulate from the front layer, and the energy separation between the symmetric and anti-symmetric wavefunctions becomes larger.
As shown in Fig.\,\ref{fig_G2K35mK} (a), $G$ drops twice at the two downward arrows. Although these two points represent two pinch-off points in the bulk 2DEGs of the front and back layers under the $\mu$m-scale gate electrodes, we assume that the calculation results above correspond to this $G$ behavior.

%%% TABLE %%%
\begin{table}%[H] add [H] placement to break table across pages
\caption{\label{Delta_SAS} Comparison of $\Delta_{\rm SAS}$ between experiment and calculation.}
\begin{ruledtabular}
\begin{tabular}{c|cc}
  &  Experiment & Calculation  \\ \hline
$\rho_{\rm S}$ ($\times 10^{10}/{\rm cm}^2$) & 6.4 & 6.2  \\
$\rho_{\rm AS}$ ($\times 10^{10}/{\rm cm}^2$) & 5.6 & 5.5  \\
$\Delta_{\rm SAS}$ (mV) & 0.29 & 0.25  \\
\end{tabular}
\end{ruledtabular}
\end{table}

\begin{acknowledgments}
We are grateful to K. Muraki and T. Saku of the NTT basic research laboratories and A. Sawada for providing us with a high mobility sample, and to M. Hashisaka and A. Ueda for their productive discussion.
This work was supported by the JSPS KAKENHI (JP15K17680, JP15H05854, JP18H01815, JP19H05826, JP19H00656).
\end{acknowledgments}

%\bibliography{QPCbib.bib}


\end{document}